\documentclass[entropy,article,submit,moreauthors,pdftex,10pt,a4paper]{mdpi} 

\usepackage{amssymb,amsmath,bm}
\usepackage{physics,cancel,nicefrac}
\firstpage{1} 
\articlenumber{x}
\doinum{10.3390/------}
\pubvolume{xx}
\pubyear{2017}
\copyrightyear{2017}
\history{Received: date; Accepted: date; Published: date}

\pdfoutput=1

\Title{Kovacs-like memory effect in athermal systems: linear
  response analysis}

\orcidauthorONE{0000-0002-4116-6854} 
\orcidauthorTWO{0000-0002-9559-8249} 

\Author{
Carlos A. Plata$^{1}$\orcidONE{}
  and Antonio
  Prados$^{1}$*\orcidTWO{}
}

\AuthorNames{Carlos A. Plata and Antonio Prados}

\address[1]{%
$^{1}$ \quad Física Teórica, Universidad de Sevilla, Apartado de
Correos 1065, E-41080 Sevilla, Spain
}

\corres{Correspondence: prados@us.es; Tel.: +34-954559514}

\abstract{We analyse the emergence of Kovacs-like memory effects in
  athermal systems within the linear response regime. This is done by
  starting from both the master equation for the probability
  distribution and the equations for the physically relevant
  moments. The general results are applied to a general class of
  models with conserved momentum and non-conserved energy. Our
  theoretical predictions, obtained within the first Sonine
  approximation, show an excellent agreement with the numerical
  results. }

\keyword{linear response; Kovacs effect; athermal system; master
  equation; H-functional  
}

\newcommand{\eq}{\text{e}}
\newcommand{\ness}{\text{s}}
\newcommand{\calP}{\mathcal{P}}
\newcommand{\calY}{\mathcal{Y}}
\newcommand{\bx}{\bm{x}}

\newcommand{\bxa}{\bm{x}_{\alpha}}
\newcommand{\bxapr}{\bm{x}_{\alpha^{\prime}}}
\newcommand{\bbW}{\mathbb{W}}

\newcommand{\vv}{\bm{v}}
\newcommand{\ra}{\rangle}
\newcommand{\la}{\langle}

\preto{\abstractkeywords}{\nolinenumbers}

\begin{document}

\section{Introduction}

The equilibrium state of physical systems is characterised by the
value of a few macroscopic variables, for example pressure, volume and
temperature in fluids. This characterisation of the equilibrium state
is \textit{complete}, in the sense that different samples sharing the
same values of the macroscopic variables respond identically to an
external perturbation. On the contrary, a system in a nonequilibrium
state, even if it is stationary, is not completely characterised by
the value of the macroscopic variables: the response to an external
perturbation depends also on additional variables or, equivalently, on
its entire thermal history. Therefore, it is often said that the
response depends on the way the system has been aged.

A pioneering work in the field of memory effects in nonequilibrium
systems was carried out by Kovacs \cite{kovacs_isobaric_1979}. The
Kovacs experiment showed that pressure, volume and temperature did not
completely characterise the state of a sample of polyvinyl acetate
that had been aged for a long time at a certain temperature
$T_{1}$. The pressure was fixed during the whole experiment and the
time evolution of the volume was recorded. At a certain time $t_{w}$,
the temperature was suddenly changed to $T$, for which the equilibrium
value of the volume equalled its instantaneous value at
$t_{w}$. Counterintuitively (thinking in equilibrium terms), the
volume did not remain constant. Instead, it displayed a hump, passing
through a maximum before tending back to its equilibrium (and initial)
value.

\begin{figure}
\begin{center}
\includegraphics[width=0.85 \textwidth]{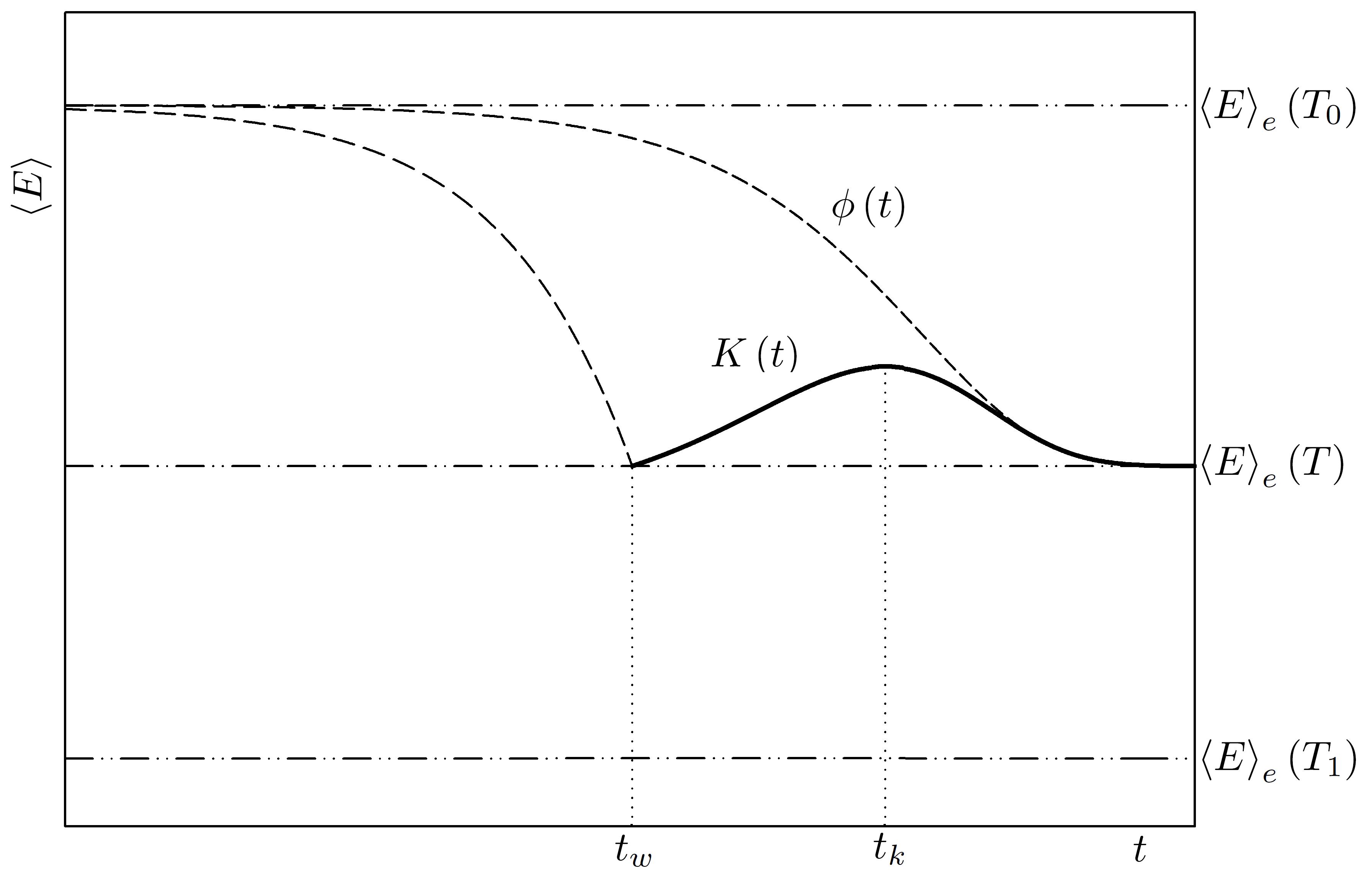}
\caption[]{Scheme of the Kovacs experiment described in the text. The
  dashed curve on the right, labelled by $\phi(t)$, represents the
  direct relaxation from $T_0$ to $T$. The dashed curve on the left
  stands for the part of the relaxation from $T_0$ to $T_1$ that is
  interrupted at $t=t_{w}$ by the second temperature jump, changing
  abruptly the temperature from $T_1$ to $T$. After this second jump,
  the system follows the non-monotonic response $K(t)$ (solid line),
  which reaches a maximum at $t=t_k$ and, afterwards, approaches
  $\phi(t)$ for very long times. \label{fig1}}
\end{center}
\end{figure}

We look into the Kovacs experiment in a more detailed way in Figure
\ref{fig1}. In recent studies of the effect in glassy systems, the
relevant physical variable is the energy instead of the volume
\cite{bertin_kovacs_2003,buhot_kovacs_2003,mossa_crossover_2004,aquino_kovacs_2006,prados_kovacs_2010,diezemann_memory_2011,ruiz-garcia_kovacs_2014}. The
system is equilibrated at a ``high'' temperature $T_0$ and at $t=0$
the temperature is suddenly quenched to a lower temperature $T$, after
which the relaxation function $\phi(t)$ of the energy $E$ is
recorded. Specifically,
$\phi(t)=\langle E(t)\rangle-\langle E\rangle_{\eq}$, where
$\langle E\rangle_{\eq}$ is the average equilibrium energy at
temperature $T$. Then, a similar procedure is followed, equilibrating
the system again at $T_{0}$, but at $t=0$ the temperature is changed
to an even lower value $T_{1}$, $T_{1}<T<T_{0}$. The system relaxes
isothermally at $T_{1}$ for a certain time $t_{w}$, such that
$\langle E\rangle(t=t_{w})$ equals $\langle E\rangle_{\eq}$. At this
time $t_{w}$, the temperature is increased to $T$ but the energy does
not remain constant: it displays the behaviour that is qualitatively
shown by $K(t)$. At first, $K(t)$ increases from zero until a maximum
is attained for $t=t_{k}$, and only afterwards it goes back to zero.
Similarly to the relaxation function, we have defined
$K(t)=\langle E(t)\rangle-\langle E\rangle_{\eq}$, for $t\geq t_{w}$.
Note that $K(t)\leq \phi(t)$ for all times, with the equality being
only asymptotically approached in the long time limit.

For molecular (thermal) systems, the equilibrium distribution has the
canonical form and it has been shown that in linear response theory
\cite{prados_kovacs_2010}

\begin{equation}\label{eq:kovacs-thermal-linear}
K(t)=\frac{T_{0}-T_{1}}{T_{0}-T} \phi(t)-\frac{T-T_{1}}{T_{0}-T}
\phi(t-t_{w}),
\end{equation}
where the final temperature $T$ and the waiting time $t_{w}$ are
related by

\begin{equation}\label{eq:T-tw-relation} \frac{T-T_{1}}{T_{0}-T_{1}}=\frac{\phi(t_{w})}{\phi(0)}.
\end{equation}
In linear response, the relaxation function $\phi(t)$ is
proportional to the equilibrium time correlation function
$\langle E(0) E(t)\rangle_{\eq}-\langle E\rangle_{\eq}^{2}$
(fluctuation-dissipation theorem), which decays monotonically in time \cite{van_kampen_stochastic_1992}.

The linear response results above make it possible to understand the
crux of the observed Kovacs hump in experiments
\cite{prados_kovacs_2010}: (i) the inequality
$0\leq K(t)\leq \phi(t)$, which assures that the hump always has a
positive sign (from now on, ``normal'' behaviour), (ii) the existence
of only one maximum in the hump and (iii) the increase of the maximum
height and the shift of its position to smaller times as $t_{w}$ is
decreased. Nevertheless, it must be noted that the experiments, both
real \cite{kovacs_isobaric_1979} and numerical
\cite{bertin_kovacs_2003,buhot_kovacs_2003,mossa_crossover_2004,aquino_kovacs_2006,diezemann_memory_2011}
are mostly done out of the linear response regime: thus, it seems that
the validity of these results extends beyond expectations. In fact, it
has been checked in simple models that the linear approximation still
gives a fair description of the hump for not-so-small temperature
jumps \cite{ruiz-garcia_kovacs_2014}.

Very recently, the investigation of Kovacs-like effects in athermal
systems has been started. A granular fluid provides a prototypical
example of an athermal system, which is intrinsically
out-of-equilibrium \cite{PL01}. A physical mechanism that inputs
energy into the system and balances in average the energy loss in
collisions, for instance the so-called stochastic thermostat
\cite{van_noije_velocity_1998}, must be considered to reach a
nonequilibrium steady state (NESS). Moreover, in general, fluctuations
are far more important in granular systems than in molecular systems
because of their smallness. The number of particles $N$, ranging
from $10^{2}$ to $10^{4}$, is large enough to make it possible to
apply the methods of statistical mechanics but definitely much smaller
than Avogadro's number.

The simplest case is that of uniformly driven granular gases
considered in \cite{prados_kovacs-like_2014,trizac_memory_2014}. The
value of the kinetic energy (granular temperature $T_{g}$) at the NESS
is controlled by the driving intensity $\xi$ of the stochastic
thermostat. Therefore, a Kovacs-like protocol can be implemented in a
completely analogous way to the one described above, with the changes
$T\to\xi$, $E\to T_{g}$.  One of the main differences found is the
emergence of ``anomalous'' Kovacs behaviour for large enough
dissipation, when $K(t)$ becomes negative and displays a minimum
instead of a maximum. It must be stressed that these results have been
obtained in the nonlinear regime, that is, for driving jumps
$\xi_{0}-\xi$, $\xi_{0}-\xi_{1}$ that are not small.

More recently, Kovacs-like behaviour has been observed in other, more
complex, athermal systems. This is the case of disordered mechanical
systems \cite{lahini_nonmonotonic_2017} and also of active matter
\cite{kursten_giant_2017}. In the latter, a ``giant'' Kovacs hump has
been reported, in the sense that the numerically observed maximum is
much larger than the one predicted by the extrapolation of the linear
approximation expression (\ref{eq:kovacs-thermal-linear}) to the
considered protocol. Moreover, an alternative derivation of
(\ref{eq:kovacs-thermal-linear}) has been provided in the supplemental
material of \cite{kursten_giant_2017}. This derivation holds for
athermal systems, since it does not make use of either the explicit
form of the probability distribution or the relationship between
response functions and time correlations at the steady state, but is
restricted to discrete-time dynamics at the macroscopic (average)
level of description.

The objectives of our paper are twofold. Firstly, we put forward a
rigorous and general derivation of the linear response expression for
the Kovacs hump for systems with a realistic continuous time
dynamics. This is done at both the mesoscopic and macroscopic levels
of description, starting from the master equation for the probability
distribution and from the hierarchy of equations for the moments,
respectively. Our proof is also valid for athermal systems, since no
hypothesis is needed with regard to the form of either the stationary
probability distribution or the fluctuation-dissipation
relation. Secondly, we apply our results to a simple class of
dissipative models that mimic the shear component of a granular fluid,
recently introduced
\cite{lasanta_fluctuating_2015,manacorda_lattice_2016,plata_lattice_2016}. Therein,
we obtain explicit expressions for the Kovacs hump within the first
Sonine approximation and compare the theoretical predictions with
numerical results. We also discuss the compatibility of the
non-monotonic behaviour displayed by the granular temperature in the
Kovacs experiment and the monotonic approach to the NESS of the
nonequilibrium entropy or $H$-functional
\cite{marconi_about_2013,de_soria_towards_2015,plata_global_2017}.

\section{Linear theory for Kovacs-like memory effects}

\subsection{General Markovian dynamics}

Here, we consider a general system, whose state is completely
characterised by a vector $\bm{x}$ with $M$ components,
$\bm{x}=\{x_{1},x_{2},\ldots,x_{M}\}$. For example, in a
one-dimensional Ising chain of $N$ spins $x_{i}=\sigma_{i}=\pm 1$, and
$M=N$; for a gas comprising $N$ particles with positions $\bm{r}_{i}$
and velocities $\bm{v}_{i}$, $M=6N$ and
$\bm{x}=\{\bm{r}_{1},\bm{v}_{1},\ldots,\bm{r}_{N},\bm{v}_{N}\}$. As is
customary and for the sake of simplicity, from now on we use a
notation suitable for systems in which the states can be labelled with
a discrete index $\alpha$, $1\leq\alpha\leq\Omega$. For example, this
is the case of the Ising system above, where $\Omega=2^{N}$. The
generalisation for a continuous index is straightforward, by changing
sums into integrals and Kronecker delta by Dirac delta
\cite{van_kampen_stochastic_1992}.

At the mesoscopic level of description, we assume that $\bm{x}$ is a
Markov process and its dynamics is governed by a master equation for
the probabilities $P(\bxa,t)$,

\begin{equation}\label{eq:master-eq}
\partial_{t}P(\bxa,t)=\sum_{\alpha^{\prime}}\left[W(\bxa|\bxapr;\xi)P(\bxapr,t)-W(\bxapr|\bxa;\xi)P(\bxa,t)\right],
\end{equation}
where $W(\bxa|\bxapr;\xi)$ are the transition rates from state $\bxapr$
to state $\bxa$, and our notation explicitly marks their dependence on
some control parameter $\xi$. Equation \eqref{eq:master-eq} can be
formally written as

\begin{equation}\label{eq:master-eq-2}
\partial_{t}\ket{\calP(t)}=\bbW(\xi)\ket{\calP(t)},
\end{equation}
where $\ket{\mathcal{P}(t)}$ is a vector (column matrix) whose components
are the probabilities $P(\bxa,t)$ and $\bbW(\xi)$ is the linear
operator (square matrix) that generates the dynamical evolution of
$\ket{\mathcal{P}(t)}$,

\begin{equation}\label{bbW}
\bbW(\bxa|\bxapr;\xi)=W(\bxa|\bxapr;\xi)-\delta_{\alpha,\alpha^{\prime}}
\sum_{\alpha^{\prime\prime}}W(\bm{x}_{\alpha^{\prime\prime}}|\bxa;\xi).
\end{equation}

Let us assume that the Markovian dynamics is ergodic (or irreducible
\cite{van_kampen_stochastic_1992}), that is, all the states are
dynamically connected through a chain of transitions with non-zero
probability. Therefore, there is a unique steady solution of the
master equation $\ket{\calP_{\ness}(\xi)}$, which verifies

\begin{equation}\label{eq:steady-sol-P}
\bbW(\xi)\ket{\calP_{\ness}(\xi)}=0,
\end{equation}
and depends on the parameter $\xi$ controlling the system
dynamics. Note that ergodicity does not imply detailed balance, so we
can have non-zero currents in the steady state. In general, this means
that the system approaches a NESS in the long time limit, not an
equilibrium state.

Now, we consider the system evolving from certain initial state at
time $t_{0}$, characterised by the distribution
$\ket{\calP(t_{0})}$. The formal solution of the master equation can
be written as

\begin{equation}\label{eq:formal-sol}
\ket{\calP(t)-\calP_{\ness}(\xi)}=e^{(t-t_{0})\bbW(\xi)}\ket{\calP(t_{0})-\calP_{\ness}(\xi)}.
\end{equation}
This is the starting point for our derivation of the expression for
the Kovacs effect in the linear response approximation, which is
carried out in the next section.

The time evolution of any physical property $Y$ is obtained right
away. Let us denote the value of $Y$ for a given configuration $\bx$
by $Y(\bx)$: its expected or average value is given by 

\begin{equation}\label{eq:average-def}
\langle Y(t)\rangle=\sum_{\alpha}Y(\bxa)P(\bxa,t)=\braket{\calY}{\calP(t)},
\end{equation}
where $\ket{\calY}$ is a ket whose components are $Y(\bxa)$, and
$\bra{\calY}$ its corresponding bra (row matrix with the same
components\footnote{We are assuming that $Y$ is a real quantity for
  all the configurations.}). By substituting \eqref{eq:formal-sol}
into \eqref{eq:average-def}, it is obtained that

\begin{equation}\label{eq:average-time-evol}
\Delta Y(t;\xi)\equiv\langle Y(t)\rangle-\langle Y\rangle_{\ness}(\xi)=\mel{\calY}{e^{(t-t_{0})\bbW(\xi)}}{\calP(t_{0})-\calP_{\ness}(\xi)},
\end{equation}
in which $\langle Y\rangle_{\ness}(\xi)$ is the average value at the
steady state of the system corresponding to $\xi$.
 
Now we investigate the relaxation of the system from the steady state
for $\xi_{0}=\xi+\Delta\xi$ to the steady state for $\xi$. We do so in
linear response, that is, $\Delta\xi$ is considered to be small and we
neglect all terms beyond those linear in $\Delta\xi$. Thus, at $t=0$
we have that the probability distribution is

\begin{equation}\label{eq:P-linear}
\ket{\calP(t=0)}=\ket{P_{\ness}(\xi+\Delta\xi)} =\ket{P_{\ness}(\xi)}+\Delta\xi \ket{\frac{dP_{\ness}(\xi)}{d\xi}}+O(\Delta\xi)^{2}.
\end{equation}
Substitution of \eqref{eq:P-linear} into \eqref{eq:average-time-evol}
yields the formal expression for the relaxation of $Y$ in linear
response,

\begin{equation}\label{eq:delta-Y-linear}
\Delta Y(t;\xi)=\Delta \xi \mel{\calY}{e^{(t-t_{0})\bbW(\xi)}}{\frac{dP_{\ness}(\xi)}{d\xi}}.
\end{equation}
In order to have an order of unity function,
one may define a normalised relaxation function

\begin{equation}\label{eq:phi-Y-linear}
\phi_{Y}(t;\xi)\equiv\lim_{\Delta \xi\to 0}\frac{\Delta
  Y(t;\xi)}{\Delta\xi}=\mel{\calY}{e^{(t-t_{0})\bbW(\xi)}}{\frac{dP_{\ness}(\xi)}{d\xi}}.
\end{equation}
Sometimes, the relaxation function is further normalised by
considering $\phi_{Y}(t)/\phi_{Y}(t=t_{0})$, see for instance
\cite{brey_stretched_1993,prados_kovacs_2010}, but such a multiplying
factor is clearly not physically relevant and will not be introduced
here.

\subsection{The Kovacs protocol. Linear response analysis from the
  master equation.}

Here, it is a Kovacs-like protocol that we introduce, by considering
that the parameter $\xi$ controlling the dynamics is changed in the
following stepwise manner:

\begin{equation}\label{eq:stepwise-xi}
\xi(t)=\left\{ 
\begin{array}{ll}
\xi_{0}, & -\infty<t<0, \\ \xi_{1}, & 0<t<t_{w}, \\ \xi, & t>t_{w}.
\end{array}\right.
\end{equation}
Therefore, since $\xi_{0}$ is kept for an infinite time, at $t=0$ the
system is prepared in the corresponding steady state,
$\calP(t=0)=\calP_{\ness}(\xi_{0})$. Our idea is to consider that the
jumps $\xi_{1}-\xi_{0}$ and $\xi-\xi_{1}$ are small, in the sense that
all expressions can be linearised in the magnitude of these
jumps. Note that this protocol is completely analogous to the Kovacs
protocol described in the introduction, see Figure \ref{fig1}, but
with $\xi$ playing the role of the temperature.

We start by analysing the relaxation in the first time window,
$0<t<t_{w}$. Therein, we apply \eqref{eq:formal-sol} with the
substitutions $t_{0}\to 0$ and $\xi\to \xi_{1}$, that is,

\begin{equation}\label{eq:evol-first-stage}
\ket{\calP(t)-\calP_{\ness}(\xi_{1})}=e^{t\bbW(\xi_{1})}
\ket{\calP_{\ness}(\xi_{0})-\calP_{\ness}(\xi_{1})}, \quad 0\leq
t\leq t_{w}.
\end{equation}
The final distribution function, at $t=t_{w}$, is the initial
condition for the next stage, $t>t_{w}$, in which the system relaxes
towards the steady state corresponding to $\xi$. Making use
again of \eqref{eq:formal-sol} with $t_{0}\to t_{w}$,

\begin{eqnarray}
\ket{\calP(t)-\calP_{\ness}(\xi)}&=& e^{(t-t_{w})\bbW(\xi)}
\ket{\calP(t_{w})-\calP_{\ness}(\xi)} \nonumber \\
& = &  e^{(t-t_{w})\bbW(\xi)} \left[ e^{t_{w}\bbW(\xi_{1})}
\ket{\calP_{\ness}(\xi_{0})-\calP_{\ness}(\xi_{1})}+ \ket{\calP_{\ness}(\xi_{1})-\calP_{\ness}(\xi)}\right]
      ,  \quad 
t\geq t_{w}.
\label{eq:evol-second-stage}
\end{eqnarray}
It must be stressed that the above expressions, Equations
\eqref{eq:evol-first-stage} and \eqref{eq:evol-second-stage}, are
exact, no approximation has been made.

The linear response approximation is introduced now: we assume that
both jumps $\xi_{0}-\xi_{1}$ and $\xi-\xi_{1}$ are small, so we can
expand both $\ket{\calP_{\ness}(\xi_{0})-\calP_{\ness}(\xi_{1})}$ and
$\ket{\calP_{\ness}(\xi_{1})-\calP_{\ness}(\xi)}$, similarly to what was
done in Equation \eqref{eq:P-linear}. Namely,

\begin{subequations}\label{eq:lin-approx}
\begin{equation}\label{eq:lin-approx-a}
\ket{\calP_{\ness}(\xi_{0})-\calP_{\ness}(\xi_{1})}=(\xi_{0}-\xi_{1})\ket{\frac{d\calP_{\ness}(\xi)}{d\xi}}+O(\xi_{0}-\xi_{1})^{2},
\end{equation}
\begin{equation}\label{eq:lin-approx-b}
\ket{\calP_{\ness}(\xi_{1})-\calP_{\ness}(\xi)}=(\xi_{1}-\xi)\ket{\frac{d\calP_{\ness}(\xi)}{d\xi}}+O(\xi_{1}-\xi)^{2}.
\end{equation}
\end{subequations}
Note that in both \eqref{eq:lin-approx-a} and \eqref{eq:lin-approx-b},
the derivatives are evaluated at $\xi$; the difference introduced by
evaluating them at either $\xi_{1}$ or $\xi_{0}$ are second order in
the deviations. Then,

\begin{equation}\label{eq:evol-second-stage-linear-1}
\ket{\calP(t)-\calP_{\ness}(\xi)}= 
(\xi_{0}-\xi_{1})  e^{(t-t_{w})\bbW(\xi)} e^{t_{w}\bbW(\xi_{1})}
\ket{\frac{d\calP_{\ness}(\xi)}{d\xi}}+
(\xi_{1}-\xi)e^{(t-t_{w})\bbW(\xi)}\ket{\frac{d\calP_{\ness}(\xi)}{d\xi}}
.
\end{equation}
This equation can be further simplified: since the two terms on its
rhs are first order in the jumps, we can substitute $\bbW(\xi_{1})$
with $\bbW(\xi)$, with the result

\begin{equation}\label{eq:evol-second-stage-linear-2}
\ket{\calP(t)-\calP_{\ness}(\xi)}= 
(\xi_{0}-\xi_{1})  e^{t\bbW(\xi)}
\ket{\frac{d\calP_{\ness}(\xi)}{d\xi}}-
(\xi-\xi_{1})e^{(t-t_{w})\bbW(\xi)}\ket{\frac{d\calP_{\ness}(\xi)}{d\xi}} .
\end{equation}
This is the superposition of two responses: the first term on the rhs
gives the relaxation from $\xi_{0}$ to $\xi_{1}$, which starts at $t=0$,
whereas the second term stands for the relaxation from $\xi_{1}$ to
$\xi$, which starts at $t=t_{w}$. We have chosen to write
$-(\xi-\xi_{1})$ in the second term because $\xi>\xi_{1}$ in the
Kovacs protocol.

The same structure in Equation \eqref{eq:evol-second-stage-linear-2}
is transferred to the average value. Taking into account Equation
\eqref{eq:average-time-evol},

\begin{equation}\label{eq:evol-Yav}
\Delta Y(t)=(\xi_{0}-\xi_{1})  \mel{\calY}{e^{t\bbW(\xi)}}
{\frac{d\calP_{\ness}(\xi)}{d\xi}}-
(\xi-\xi_{1})\mel{\calY}{e^{(t-t_{w})\bbW(\xi)}}{\frac{d\calP_{\ness}(\xi)}{d\xi}},
\quad t\geq t_{w},
\end{equation}
in which we recognise the relaxation function in linear response,
defined in Equation \eqref{eq:phi-Y-linear}. We can also normalise the
response in this experiment, by defining a function $K(t)$
as follows, 

\begin{equation}\label{eq:evol-Yav-phi}
K_{Y}(t)\equiv \lim_{\xi_{0}\to\xi}\frac{\Delta Y(t)}{\xi_{0}-\xi}=\frac{\xi_{0}-\xi_{1}}{\xi_{0}-\xi}  \phi_{Y}(t)-
\frac{\xi-\xi_{1}}{\xi_{0}-\xi} \phi_{Y}(t-t_{w}).
\end{equation}
It is understood that, as $\xi_{0}-\xi\to 0$, both
prefactors $\frac{\xi_{0}-\xi_{1}}{\xi_{0}-\xi}$ and
$\frac{\xi-\xi_{1}}{\xi_{0}-\xi}$ are kept of the order of unity.

A few comments on Equation \eqref{eq:evol-Yav-phi} are in
order. Hitherto, no restriction has been imposed on the state of the
system at $t=t_{w}$; therefore, \eqref{eq:evol-Yav-phi} is valid for
arbitrary $(\xi_{0},\xi_{1},\xi)$, provided that the jumps are small
enough and the ratio of the jumps is of the order of unity. The
function $K(t)$ corresponds to a Kovacs-like experiment when $\xi$ is
chosen as a function of $t_{w}$ in such a way that $\langle
Y(t_{w})\rangle=\langle Y\rangle_{\ness}(\xi)$ or $K_{Y}(t_{w})=0$, that is,

\begin{equation}\label{eq:Kovacs-condition}
 \frac{\xi-\xi_{1}}{\xi_{0}-\xi_{1}}=\frac{\phi_{Y}(t_{w})}{\phi_{Y}(0)}.
\end{equation} 
Alternatively, one may consider that \eqref{eq:Kovacs-condition}
defines $t_{w}$ as a function of $\xi$. 

The complete analogy between Equations \eqref{eq:evol-Yav-phi} and
\eqref{eq:Kovacs-condition} and Equations
\eqref{eq:kovacs-thermal-linear} and \eqref{eq:T-tw-relation} is
apparent. Nevertheless, we have made use neither of the explicit form
of the steady state distribution (in general non-canonical) nor of the
relation between response functions and time correlation functions
(fluctuation-dissipation relation), which were necessary in
\cite{prados_kovacs_2010} to demonstrate
Equation~\eqref{eq:kovacs-thermal-linear}. Therefore, the proof
developed here is more general, being valid for any steady state,
equilibrium or nonequilibrium, and thus it specifically holds in
athermal systems. Also, it must be noted that it can be easily
extended to the Fokker-Planck, or the equivalent Langevin, equation.

\subsection{Linear response from the equations for the moments}

In this section, we do not start from the equation for the probability
distribution, but from the equations for the relevant physical
properties of the considered system. For example, one may think of the
hydrodynamic equations for a fluid or the law of mass action equations
for chemical reactions. Of course, these equations can be derived in a
certain ``macroscopic'' approximation
\cite{van_kampen_stochastic_1992}, which typically involves neglecting
fluctuations, from the equation for the probability distribution by
taking moments, but this is not our approach here. Anyhow, we borrow
this term to call our starting point ``equations for the moments''.

We denote the relevant moments by $z_{i}$, $i=1,\ldots,J$,
where $J$ is the number of relevant moments. The equations for the moments have the general form

\begin{equation}\label{eq:moments-eq}
\frac{d}{dt}z_{i}=f_{i}(z_{1},\ldots,z_{J};\xi),
\end{equation}
where $f_{i}$ are continuous, in general nonlinear, functions of the
moments. This is a key difference between moment equations and the
master (or Fokker-Planck) equation, since the latter is always linear
in the probability distribution. Therefore, unlike the master
equation, Equation \eqref{eq:moments-eq} cannot be formally solved for
arbitrary initial conditions. Notwithstanding, in the linear response
approximation, we show here that a procedure similar to the one
carried out in the previous section leads to the same expression for
the Kovacs hump.

We assume that there is only one steady solution of Equation
\eqref{eq:moments-eq} that is globally stable, at which the
corresponding values of the moments are
$z_{i}^{\ness}(\xi)$. Linearisation of the dynamical system around the
steady state gives

\begin{equation}\label{eq:lin-dyn-syst}
\frac{d}{dt} \ket{\Delta z(t)} =\mathbb{M}(\xi) \ket{\Delta z(t)}, \qquad
\ket{\Delta z(t)}\equiv \ket{z(t)-z_{\ness}(\xi)}.
\end{equation}
We are using a notation completely similar to that in the previous
section, $\ket{z}$ is a vector, represented by a column matrix with
components $z_{i}$, and $\mathbb{M}(\xi)$ is a linear operator,
represented by a square matrix with elements

\begin{equation}\label{eq:M-matrix-elem}
 M_{ij}(\xi)=\left. \partial_{z_{j}}f_{i} \right|_{\ket{z}=\ket{z_{\ness}(\xi)}}.
\end{equation}
Note that the dimensions of these matrices are much smaller than those
for the master equation, since $J$ is of the order of unity and does
not diverge in the thermodynamic limit. In general,
$M_{ij} \neq M_{ji}$ and the operator $\mathbb{M}$ is not
Hermitian. However, we do not need $\mathbb{M}$ to be Hermitian to
solve the linearised system in a formal way, as shown below.

Analogously to what was done for the master equation, the formal
solution of Equation \eqref{eq:lin-dyn-syst} is

\begin{equation}\label{eq:lin-dyn-syst-sol1}
\ket{\Delta z(t)}=e^{(t-t_{0})\mathbb{M}(\xi)}\ket{\Delta z(t_{0})}.
\end{equation}
In particular, if the initial condition is chosen to
correspond to the steady state for $\xi_{0}=\xi+\Delta\xi$, one has

\begin{equation}\label{eq:lin-dyn-syst-sol2}
\ket{\Delta z(t)}=\Delta\xi\, e^{(t-t_{0})\mathbb{M}(\xi)}\ket{\frac{dz_{\ness}(\xi)}{d\xi}}.
\end{equation}
The response for any of the relevant moments can be extracted by
projecting the above result onto the ``natural'' basis
$\ket{u_{i}}$, whose components are $u_{ij}=\delta_{ij}$. Then, the
normalised linear response function for $z_{i}$ can be defined by

\begin{equation}\label{eq:lin-resp-normalised}
  \phi_{z_{i}}(t)=\lim_{\Delta\xi\to 0}\frac{\braket{u_{i}}{\Delta z(t)}}{\Delta\xi}= \mel{u_{i}}{e^{(t-t_{0})\mathbb{M}(\xi)}}{\frac{dz_{\ness}(\xi)}{d\xi}}.
\end{equation}
Note the utter formal analogy of expression \eqref{eq:lin-resp-normalised}  with Equation \eqref{eq:phi-Y-linear},
which was obtained from the master equation. The proof of the
expression for the Kovacs hump follows exactly the same line of
reasoning, and the result is exactly that in Equations
\eqref{eq:evol-Yav-phi} and \eqref{eq:Kovacs-condition}, and thus it
is not repeated here.

\section{A lattice model with conserved momentum and non-conserved
  energy}
\label{Section-model}
\subsection{Definition of the model. Kinetic description}

Here, we briefly put forward a class of models for the shear component
of the velocity in a granular gas, focusing on the features that are
needed for the discussion of memory effects. A more complete
description of the model, including its physical motivation, can be
found in
\cite{lasanta_fluctuating_2015,manacorda_lattice_2016,plata_lattice_2016,prasad_driven_2016}.

The system is defined on a $1$d lattice: there is a particle with
velocity $v_{l}$ at each site $l$. The system configuration is thus
given by $\vv \equiv \{v_{1},...,v_{N}\}$. The dynamics proceeds
through inelastic nearest-neighbour binary collisions: each pair
$(l,l+1)$ collides inelastically with a characteristic rate
proportional to $|v_{l}-v_{l+1}|^{\beta}$, with $\beta\geq 0$. For
$\beta=0$, nearest-neighbour particles collide independently of their
relative velocity (the so-called Maxwell-molecule model \cite{BK03}),
whereas for $\beta=1$ and $\beta=2$ we have a collision rate analogous
to that of hard spheres and very hard spheres, respectively
\cite{PL01,ETB06a}. The pre-collisional velocities are transformed
into the post-collisional ones by the operator $\hat{b}_{l}$,

\begin{subequations}\label{eq:coll_rule}
\begin{eqnarray}
\hat{b}_{l}v_{l} &=& v_{l}-\frac{1+\alpha}{2}\left(v_{l}-v_{l+1}\right), \\
\hat{b}_{l}v_{l+1} &=& v_{l+1}+\frac{1+\alpha}{2}\left(v_{l}-v_{l+1}\right),
\end{eqnarray}
\end{subequations}
where $0 < \alpha\leq 1$ is the normal restitution
coefficient. Momentum is always conserved in collisions,
$(\hat{b}_{l}-1)(v_{l}+v_{l+1})=0$, but energy is not, in fact
$(\hat{b}_{l}-1)(v_{l}^{2}+v_{l+1}^{2}) =
(\alpha^{2}-1)(v_{l}-v_{l+1})^{2}/2\leq 0$,
with equality holding only in the elastic limit $\alpha=1$.

In order to have a steady state, a mechanism that injects energy into
the system, and thus compensates in average the energy loss in
collisions, must be introduced. For the sake of simplicity, we
consider here that the system is ``heated'' by a white noise force,
that is, the so-called stochastic thermostat
\cite{van_noije_velocity_1998,MS00}. The velocity change introduced by
the stochastic forcing is

\begin{equation}\label{jump-moments-1}
  \left.\partial_{\tau} v_{i}(\tau)\right|_{\text{noise}} = \xi_{i}(\tau)-\frac{1}{N}\sum_{j=1}^{N}
    \xi_{j}(\tau), 
\end{equation}
and $\xi_{i}(\tau)$ are Gaussian
white noises of amplitude $\chi$

\begin{equation}\label{jump-moments-2}
\la \xi_{i}(\tau)\ra_{\text{noise}}=0, \quad \la \xi_{i}(\tau)\xi_{j}(\tau')\ra_{\text{noise}}=\chi
\delta_{ij}\delta(\tau-\tau'),
\end{equation}
for $i,j=1,\ldots,N$. This version of the stochastic thermostat
conserves total momentum, which is needed to assure the existence of a
well-defined steady state
\cite{maynar_fluctuating_2009,prasad_high-energy_2013}.

We do not write here the master-Fokker-Planck equation for the
$N$-particle $P_{N}(\vv,\tau)$ probability density of finding the
system in state $\vv$ at time $\tau$. Instead, we directly write the
``kinetic'' equation for the one-particle distribution function,
namely $ P_{1}(v;l,\tau)=\int d\vv P_{N}(\vv,\tau) \delta(v_{l}-v)$.
Therefrom, all the one-site velocity moments, which are the relevant
physical quantities for our present purposes, can be derived,
$\la v_{l}^{n}(\tau)\ra\equiv \int_{-\infty}^{+\infty}dv\, v^{n}
P_{1}(v;l,\tau)$.
As usual in kinetic theory, a closed equation for $P_{1}$ can be
written only after introducing the Molecular Chaos assumption: spatial
correlations at different sites are of the order of $N^{-1}$ and
then negligible.

We analyse here homogeneous states; if the initial state is
homogeneous, the above dynamics preserves homogeneity. Moreover, we
employ the usual notation in kinetic theory,
$f(v,\tau)\equiv P_{1}(v,\cancel{l};\tau)$ and consider the
quasi-elastic limit, that is, $\epsilon\equiv 1-\alpha^{2}\ll 1$. This
allows us to write a simpler expression for the inelastic collision
term. Proceeding along the same lines as in
\cite{manacorda_lattice_2016,plata_global_2017}, it is obtained that \cite{PlataPradosunpub17}

\begin{equation}\label{eq:kinetic-eq-tau}
\partial_{\tau}f(v,\tau)=\frac{\omega\epsilon}{2}\partial_{v}\int_{-\infty}^{+\infty}dv^{\prime}
(v-v^{\prime}) |v-v^{\prime}|^{\beta}
f(v,\tau)f(v^{\prime},\tau)+\frac{\chi}{2} \partial_{v}^{2} f(v,\tau).
\end{equation}
We can define a dimensionless time scale, $t=\omega\epsilon\tau$, over
which 

\begin{equation}\label{eq:kinetic-eq-t}
\partial_{t}f(v,t)=\frac{1}{2}\partial_{v}\int_{-\infty}^{+\infty}dv^{\prime}
(v-v^{\prime}) |v-v^{\prime}|^{\beta}
f(v,t)f(v^{\prime},t)+\frac{\xi}{2} \partial_{v}^{2} f(v,t),
\end{equation}
where $\xi$ is the rescaled strength of the noise,
$\xi=\frac{\chi}{\omega\epsilon}$. Note that this kinetic equation is,
like Boltzmann's or Enskog's, nonlinear in $f(v,t)$.

The main physical magnitude is the granular temperature $T$, which we
define as

\begin{equation}\label{eq:granular-T}
T\equiv \langle v^{2}\rangle=\int_{-\infty}^{+\infty}dv\, v^{2} f(v,t).
\end{equation}
We used the notation $T_{g}$ for the granular temperature in
the introduction, to differentiate it from the usual thermodynamic
temperature $T$.  Since the latter plays no role in our system, we
employ the usual notation $T$ for the granular temperature hereafter.
It is also customary to define the thermal velocity $v_{0}=\sqrt{2T}$
to make $v$ dimensionless, with the change

\begin{equation}\label{eq:v-dimensionless}
v=v_{0}c, \qquad f(v,t)dv=\varphi(c,t)dc \Leftrightarrow \varphi(c,t)=v_{0}f(v,t).
\end{equation}
Taking moments in \eqref{eq:kinetic-eq-t} and making the change of
variables above, one gets

\begin{equation}\label{eq:evol-T-1}
\frac{d}{d t}T=-\zeta\,T^{1+\frac{\beta}{2}}+\xi, \qquad \zeta=2^{\frac{\beta}{2}}\int_{-\infty}^{+\infty} dc
\int_{-\infty}^{+\infty} dc^{\prime}
|c-c^{\prime}|^{2+\beta}\varphi(c,t)\varphi(c^{\prime},t)
\end{equation}
The first term on the rhs stems from collisions and \textit{cools} the
system, in the sense that it always make the granular temperature
decrease. The second term stems from the stochastic thermostat and
\textit{heats} the system and, thus, in the long time limit a NESS is
attained in which both terms counterbalance each other.

\subsection{First Sonine approximation}

The equation for the granular temperature is not closed in general and
then an expansion in Sonine (or Laguerre) polynomials is typically
introduced,

\begin{equation}\label{eq:Sonine-expansion}
\varphi(c,t)=\frac{e^{-c^{2}}}{\sqrt{\pi}}\left[
  1+\sum_{k=2}^{\infty}a_{k}(t) L_{k}^{\left(-\frac{1}{2}\right)}(c^{2})\right],
\end{equation}
where $L_{k}^{(m)}(x)$ are the associated Laguerre polynomials
\cite{AS72}. In kinetic theory, $m=\frac{d}{2}-1$, with $d$ being the
spatial dimension, and often the notation
$S_{k}(x)\equiv L_{k}^{\left(\frac{d}{2}-1\right)}$ is used. Here, we
mainly use the so-called first Sonine approximation, in which (i) only
the term with $k=2$ is retained and (ii) nonlinear terms in $a_{2}$
are neglected. The coefficient $a_{2}$ is the excess kurtosis,
$\langle c^{4}\rangle=3(1+a_{2})/4$.

Although the linearisation in $a_{2}$ is quite standard in kinetic
theory, we derive firstly the evolution equations considering just
step (i) of the first Sonine approximation, that is, we truncate the
expansion for the scaled distribution \eqref{eq:Sonine-expansion}
after the $k=2$ term. Henceforth, we call this approximation,
nonlinear first Sonine approximation. Afterwards, in the numerical
results, we will discuss how both approximations, nonlinear and
standard, give almost indistinguishable results.

In the nonlinear first Sonine approximation, it is readily obtained
 the evolution equation for the temperature \cite{PlataPradosunpub17}
 
\begin{subequations}\label{eq:evol-Soninenl}
\begin{equation}\label{eq:evol-T-Soninenl}
\frac{d}{dt}T=- \zeta_{0}\,
T^{1+\frac{\beta}{2}}\left[1+\frac{\beta(2+\beta)}{16}a_{2}+\frac{\beta(2+\beta)(2-\beta)(4-\beta)}{1024}a_{2}^{2}\right]+\xi,
\end{equation}
where
$\zeta_{0}=\pi^{-\nicefrac{1}{2}}\,2^{1+\beta}\,\Gamma\!\left(\frac{3+\beta}{2}\right)$.
Unless $\beta=0$ (Maxwell molecules), the equation for the temperature
is not closed. Then, we write down the equation for $a_{2}$: again,
after a lengthy but straightforward calculation we derive

\begin{align}
T\frac{d}{dt}a_{2}=&-2\xi a_{2}- \frac{\zeta_{0}}{3}
  \beta\, T^{1+\frac{\beta}{2}}  \nonumber \\
&  \times
\left[1+\frac{56+\beta(6+\beta)}{16}a_{2}-\frac{(2+\beta)[384+(2-\beta)\beta(4+\beta)]}{1024}a_{2}^{2}-\frac{3(4-\beta)(2-\beta)(2+\beta)}{512}
a_{2}^{3}\right].
  \label{eq:evol-a2-Soninenl}
\end{align}
\end{subequations}
The evolution equations in the standard first Sonine approximation are
easily reached just neglecting nonlinear terms in $a_2$ in Equations
\eqref{eq:evol-Soninenl}, that is,

\begin{subequations}\label{eq:evol-Sonine}
\begin{equation}\label{eq:evol-T-Sonine}
\frac{d}{dt}T=- \zeta_{0}\,
T^{1+\frac{\beta}{2}}\left[1+\frac{\beta(2+\beta)}{16}a_{2}\right]+\xi,
\end{equation}
\begin{equation}\label{eq:evol-a2-Sonine}
  T\frac{d}{dt}a_{2}=- \frac{\zeta_{0}}{3}
  \beta\, T^{1+\frac{\beta}{2}}
  \left[1+\frac{56+\beta(6+\beta)}{16}a_{2}\right]-2\xi a_{2}.
\end{equation}
\end{subequations}
For $\xi\neq 0$, the steady solution of these equations is

\begin{equation}\label{eq:T-a2-steady}
 T_{s}=\left(\frac{\xi}{\zeta_{0}\left[1+
\frac{\beta(2+\beta)}{16}a_{2}^{\ness}\right]}\right)^{\frac{2}{2+\beta}},\qquad a_{2}^{\ness}=-\frac{16\beta}{96+56\beta+6\beta^{2}+\beta^{3}}.
\end{equation}
Note that (i) $0\leq |a_{2}^{\ness}|\leq 0.133$ for
$0\leq \beta\leq 2$, which makes it reasonable to use the first Sonine
approximation, and (ii) $a_{2}^{\ness}$ is independent of the driving
intensity $\xi$. This will be useful in the linear
response analysis, to be developed below, because a sudden change in
the driving only changes the stationary value of the temperature but
not that of the excess kurtosis. If $\xi=0$, the system evolves
towards the homogeneous cooling state, in which the excess kurtosis
tends to the value

\begin{equation}
\label{eq:a2HCS}
a_{2}^{\text{HCS}}=-\frac{16}{56+\beta(6+\beta)},
\end{equation}
as predicted by Equation \eqref{eq:evol-a2-Sonine}, and the
temperature decays  following Haff's law, $dT/dt \propto -T^{1+\frac{\beta}{2}}$.

From now on, we use reduced temperature and time,

\begin{equation}\label{eq:theta-A2-def}
\theta=\frac{T}{T_{\ness}}, \quad s=\zeta_{0}T_{s}^{\beta/2}t.
\end{equation}
The steady temperature $T_{\ness}$ plays the role of a natural energy
(or granular temperature) unit. In reduced variables, the evolution
equations are

\begin{subequations}\label{eq:reduced-var-evol}
\begin{eqnarray}\label{eq:reduced-var-evol-1}
  \frac{d}{ds}\theta&=&1-\theta^{1+\frac{\beta}{2}}+\frac{\beta(2+\beta)}{16}
                        \left(a_{2}^{\ness}-a_{2}\theta^{1+\frac{\beta}{2}}\right), \\
  \theta \frac{d}{ds}a_{2}&=& \kappa_{1} \left( a_{2} - a_{2}^{\text{HCS}} \right) \left( 1-\theta^{1+\frac{\beta}{2}} \right) -\kappa_{2} \left(a_{2}-a_{2}^{\ness}\right),
\label{eq:reduced-var-evol-2}
\end{eqnarray}
\end{subequations}
where we have introduced two parameters of the order of unity, 

\begin{equation}\label{eq:B-def}
\kappa_{1}=-\frac{\beta}{3 a_{2}^{\text{HCS}}}, \quad \kappa_{2}=-\frac{\beta}{3 a_{2}^{\ness}},
\end{equation}
$0\leq \kappa_{1}\leq 3$ and $2\leq \kappa_{2}\leq 5$ for
$0\leq\beta\leq 2$. 

The evolution equations in the first Sonine approximation,
\eqref{eq:evol-Sonine} or \eqref{eq:reduced-var-evol}, are the
particularisation of the equations for the moments
\eqref{eq:moments-eq} to our model: $J=2$, and $z_{1}=T$ (or
$\theta$), $z_{2}=a_{2}$. Consistently, they are nonlinear although
here, due to the simplifications introduced in the first Sonine
approximation, only nonlinear in $\theta$. When the system is close to
the NESS, Equations \eqref{eq:reduced-var-evol} can be linearised
around it by writing $\theta=1+\Delta\theta$,
$a_{2}=a_{2}^{\ness}+\Delta a_{2}$,

\begin{equation}\label{eq:linearised-system}
  \frac{d}{ds}\begin{pmatrix}\Delta\theta \\ \Delta a_{2} \end{pmatrix}
  =\bm{M}\cdot\begin{pmatrix}\Delta\theta \\ \Delta a_{2} \end{pmatrix},
  \qquad
  \bm{M}=\begin{pmatrix} -\frac{2(2+\beta)(12+\beta)}{48+4\beta+\beta^{2}} &
    -\frac{\beta(2+\beta)}{16} \\ \kappa_{1} \left( 1 + \frac{\beta}{2}  \right) \left( a_{2}^{\text{HCS}}-a_{2}^{\ness} \right)  & -\kappa_{2}
 \end{pmatrix} .
\end{equation}
Of course, the general solution of this linear system for
arbitrary initial conditions $\Delta\theta(0)$ and $\Delta a_{2}(0)$
can be immediately written, but we omit it here.

\subsection{Kovacs hump in linear response}

Now we look into the Kovacs hump in the linear response
approximation. Following the discussion leading to Equation
\eqref{eq:evol-Yav-phi}, first we have to calculate the relaxation
function $\phi_{T}$ for the granular temperature. The system is at the
steady state corresponding to a driving $\xi_{0}$ for
$t<0$, at $t=0$ the driving is instantaneously changed to $\xi$ and
only the linear terms in $\Delta\xi=\xi-\xi_{0}$ are retained. We
choose the normalisation of $\phi_{T}(s)$ in such a way that
$\phi_{T}(0)=1$, that is

\begin{equation}\label{phi-granular-temp-def}
\phi_{T}(s)\equiv\lim_{\Delta T(0)\to 0}\frac{\Delta T(s)}{\Delta
  T(0)}=\lim_{\Delta\theta(0)\to 0}\frac{\Delta\theta(s)}{\Delta\theta(0)}.
\end{equation} 

Since $T_{\ness}$ changes with $\xi$ but $a_{2}$ does not, we have to
solve Equation~\eqref{eq:linearised-system} for $\Delta a_{2}(0)=0$
and arbitrary (small enough) $\Delta\theta(0)$. The solution is

\begin{subequations}\label{phi-granular-temp-exp}
\begin{equation}
\phi_{T}(s)=c_{+}e^{\lambda_{+}s}+c_{-}e^{\lambda_{-}s},
\end{equation}
\begin{equation}
c_{+}=\frac{M_{11}-\lambda_{-}}{\lambda_{+}-\lambda_{-}}, \qquad
c_{-}=\frac{\lambda_{+}-M_{11}}{\lambda_{+}-\lambda_{-}},
\end{equation}
\end{subequations}
where $M_{ij}$ is the $(i,j)$ element of the matrix $\bm{M}$ and
$\lambda_{\pm}$ its eigenvalues,

\begin{equation}\label{eq:eigenv}
\lambda_{\pm}=\frac{\Tr\bm{M}\pm\sqrt{(\Tr\bm{M})^{2}-4\det\bm{M}}}{2}=\frac{\Tr\bm{M}\pm\sqrt{(M_{11}-M_{22})^{2}+4M_{12}M_{21}}}{2}.
\end{equation}
Both eigenvalues $\lambda_{\pm}$ are negative, since $\Tr\bm{M}<0$ and
$\det\bm{M}>0$ for all $\beta>0$. Therefore,
$|\lambda_{+}|<|\lambda_{-}|$ and it is $\lambda_{+}$ that dominates
the relaxation of the granular temperature for long times.  Moreover,
$c_{\pm}>0$ and thus the linear relaxation function $\phi_{T}(s)$ is
always positive and decays monotonically to zero.

Next we consider a Kovacs-like experiment: the system was at the NESS
corresponding to a driving $\xi_{0}$, with granular temperature
$T_{\ness,0}$ for $t<0$, the driving is suddenly changed to $\xi_{1}$
at $t=0$ so that the system starts to relax towards a new steady
temperature $T_{\ness,1}$ for $0\leq t\leq t_{w}$, and this relaxation
is interrupted at $t=t_{w}$, because the driving is again suddenly
changed to the value $\xi$ such that the stationary granular
temperature $T_{\ness}$ equals its instantaneous value at $t_{w}$. The
time evolution of the granular temperature for $t\geq t_{w}$ is given
by the particularisation of Equations \eqref{eq:evol-Yav-phi} and
\eqref{eq:Kovacs-condition} to our situation, that is,

\begin{equation}\label{eq:Kovacs-Sonine}
K_{T}(s)=\frac{\xi_{0}-\xi_{1}}{\xi_{0}-\xi}  \phi_{T}(s)-
\frac{\xi-\xi_{1}}{\xi_{0}-\xi} \phi_{T}(s-s_{w}), \qquad
 \frac{\xi-\xi_{1}}{\xi_{0}-\xi_{1}}=\phi_{T}(s_{w}),
\end{equation} 
where we have made use of the normalisation $\phi_{T}(0)=1$. In the
linear response approximation, the jumps in the driving values can be
substituted by the corresponding jumps in the stationary values of the
granular temperature.

It is important to stress that the positivity and monotonic decay to
zero of the linear relaxation function $\phi_{T}(s)$ assures that the
Kovacs behaviour is \textit{normal}, that is, (i) $K_{T}(s)$ is always
positive and bounded from above by $\phi_{T}(s)$ and (ii) there is
only one maximum at a certain time $s_{k}>s_{w}$
\cite{prados_kovacs_2010}. The \textit{anomalous} behaviour found in
the uniformly heated hard-sphere granular for large enough
inelasticity  is thus not present here. This is consistent with the  
quasi-elastic limit we have introduced to simplify the collision
operator.

\subsection{Nonlinear Kovacs hump}

Here we consider the Kovacs hump for arbitrary large driving jumps. In
our model, we can make use of the smallness of $a_{2}$, which is
assumed in the first Sonine approximation, in order to introduce a
perturbative expansion of Equations \eqref{eq:reduced-var-evol} in powers
of $a_{2}^{\ness}$. The procedure is completely analogous to that
performed in \cite{prados_kovacs-like_2014,trizac_memory_2014} for a
dilute gas of inelastic hard spheres and thus we omit the details
here. 

We start by writing $a_{2}=a_{2}^{\ness}A_{2}$, with $A_{2}$ of the
order of unity, and

\begin{equation}\label{eq:generic-expansion-in-a2}
\theta(s)=\theta_{0}(s)+a_{2}^{\ness}\theta_{1}(s)+\ldots, \qquad A_{2}(s)=A_{20}(s)+a_{2}^{\ness}A_{21}(s)+\ldots.
\end{equation}
These expansions are inserted into Equations
\eqref{eq:reduced-var-evol}, which have to be solved with initial
conditions $\theta(s_{w})=1$, $A_{2}(s_{w})=A_{2}^{\text{ini}}$. To
the lowest order, $\theta_{0}(s)=1$ whereas $A_{20}(s)$ decays
exponentially to one,

\begin{equation}\label{eq:expansion_a2s-kurtosis}
 A_{20}(s) - 1 \sim \left( A_{2}^{\text{ini}} - 1 \right) e^{-\kappa_{2} \left(s-s_{w} \right)}.
\end{equation} 
In order to describe the Kovacs hump, we compute 
$\theta_{1}(s)$ that verifies the differential equation

\begin{equation}\label{eq:theta1-evol}
\frac{d\theta_{1}}{ds}=-\left(1+\frac{\beta}{2}\right)\theta_{1}+\frac{\beta(2+\beta)}{16}\left(A_{20}-1\right),
\end{equation}
which can be immediately integrated to give 

\begin{equation}\label{eq:expansion_a2s-theta}
\theta(s)-1 \sim \left(a_{2}^{\text{ini}}-a_{2}^{\ness}\right)
\frac{ \beta (2+\beta)}{8(2+\beta-2\kappa_{2})} \left[
e^{-\kappa_{2}\left(s-s_{w} \right)}-e^{-\left(1+\frac{\beta}{2}
  \right) \left(s-s_{w} \right)} \right], \qquad s\geq s_{w},
\end{equation}
The structure of this result is completely analogous to those in
\cite{prados_kovacs-like_2014,trizac_memory_2014} and thus the
conclusions can also be drawn in a similar way. In particular, we want
to highlight that (i) the factor that controls the size of the hump is
proportional to $a_{2}^{\text{ini}}-a_{2}^{\ness}$, and (ii) the
shape of the hump is codified in the factor between brackets that only
depends on $\beta$. Note that $(a_{2}^{\text{ini}}-a_{2}^{\ness})>0$
for the \textit{cooling} protocols ($\xi_{1}<\xi<\xi_{0}$) considered
here and thus no anomalous Kovacs hump is expected in the nonlinear
regime either.

\section{Numerical results}

Here, we compare the theoretical approach above to numerical results
of our model. Specifically, we focus on the case $\beta=1$ that gives
a collision rate similar to that of hard-spheres. All simulations have
been carried out with a restitution coefficient $\alpha=0.999$, which
corresponds to the quasi-elastic limit in which our simplified kinetic
description holds. Also, we set $\omega=1$ without loss of generality.

\subsection{Validation of the evolution equations and linear relaxation}

First of all, we check the validity of the first Sonine approximation,
as given by Equations \eqref{eq:evol-Sonine}, to describe the time
evolution of our system. In order to do so, we compare several
relaxation curves between the NESS corresponding to two different
noise strengths. In particular, we always depart from the stationary
state corresponding to $\chi_{0}=1$ and afterwards let the system
evolve with $\chi=\{ 0.2, 0.6, 0.8, 1 \}$ for $t>0$. In Figure
\ref{fig2}, we compare the Monte-Carlo simulation (see appendix
\ref{appA} for details) of the system with the numerical solution of
the evolution equations in the first Sonine approximation
\eqref{eq:evol-Sonine}. In addition,
we also have plotted the analytical solution of the linear response
system, Equation \eqref{eq:linearised-system}.  The agreement is
complete between simulation and theory, and it can be observed how the
linear response result becomes more accurate as the temperature jump
decreases.

\begin{figure}
\begin{center}
\includegraphics[width=0.7 \textwidth]{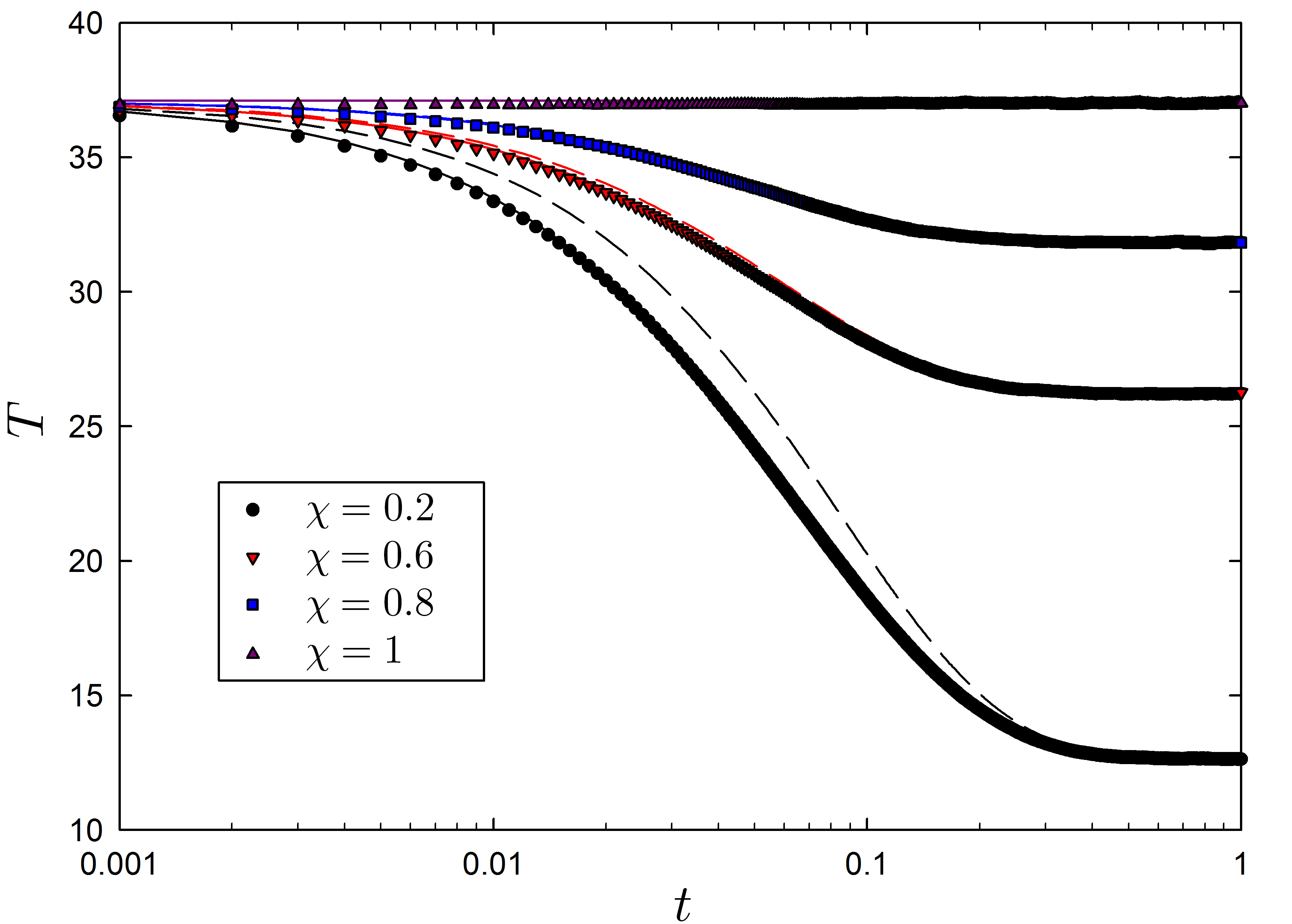}
\caption[]{(Color online) Direct relaxation of the granular
  temperature $T$ for different final noise amplitudes. All curves
  start from the stationary state corresponding to $\chi_{0}=1$. We
  compare Monte-Carlo simulation results for a system of $N=100$ sites
  (symbols) with the numerical solution of the first Sonine
  approximation, Equations \eqref{eq:evol-Sonine} (solid lines), and
  the analytic solution of the linear response system, Equation
  \eqref{eq:linearised-system} (dashed lines). \label{fig2}}
\end{center}
\end{figure}

In order not to clutter the plot in Figure \ref{fig2}, we have not
shown the results for the nonlinear first Sonine approximation,
Equations \eqref{eq:evol-Soninenl}. The relative error between their
numerical solution and that of the standard first Sonine approximation
\eqref{eq:evol-Sonine} is at most of order $0.1 \%$, for all the cases
we have considered. Henceforth, we always use the latter, which is the
usual approach in kinetic theory.

\subsection{Kovacs hump}

Since the numerical integration of the first Sonine approximation
perfectly agrees with Monte Carlo simulations, we compare the
analytical results for the Kovacs hump with the former.  Specifically,
we work in reduced variables and therefore we integrate numerically
Equations \eqref{eq:reduced-var-evol}.

\subsubsection{Linear response}

It is convenient to rewrite the expression for the Kovacs hump in an
alternative form to compare our theory to numerical results. We take
advantage of the simple structure of the relaxation function in the
first Sonine approximation, which is the sum of two exponentials, to
introduce the factorisation \cite{prados_kovacs_2010}

\begin{subequations}\label{eq:Kovacs-factorisation-and-factors}
\begin{equation}\label{eq:Kovacs-factorisation}
K_{T}(s)=K_{0}(s_{w})K_{1}(s-s_{w}),
\end{equation}
where

\begin{equation}\label{eq:Kovacs-factors}
K_{0}(s_{w})=c_{+}c_{-}
\frac{e^{\lambda_{+}s_{w}}-e^{\lambda_{-}s_{w}}}{1-\phi_{T}(s_{w})},
\qquad
K_{1}(s-s_{w})=e^{\lambda_{+}(s-s_{w})}-e^{\lambda_{-}(s-s_{w})}.
\end{equation}
\end{subequations}

Firstly, this factorisation property shows that the position $s_{k}$
of the maximum relative to the waiting time $s_{w}$, that is,
$s_{k}-s_{w}$, is controlled by the function $K_{1}$. Thus,
$s_{k}-s_{w}$ does not depend on the waiting time but only on the two
eigenvalues $\lambda_{\pm}$.  Namely,

\begin{equation}\label{eq:max-pos}
s_{k}-s_{w}=\frac{1}{\lambda_{+}-\lambda_{-}}
\ln\left(\frac{\lambda_{-}}{\lambda_{+}}\right) \underset{\beta=1}{\simeq} 0.442.
\end{equation}
Secondly, the height of the maximum $K_{\max}$ does depend on the
waiting time $s_{w}$ due to the factor $K_{0}(s_{w})$. Specifically,
it can be shown that $K_{\max}$ is a monotonically decreasing function
of the waiting time $s_{w}$ that vanishes in the limit as
$s_{w}\to\infty$.

\begin{figure}
\begin{center}
\includegraphics[width=0.7 \textwidth]{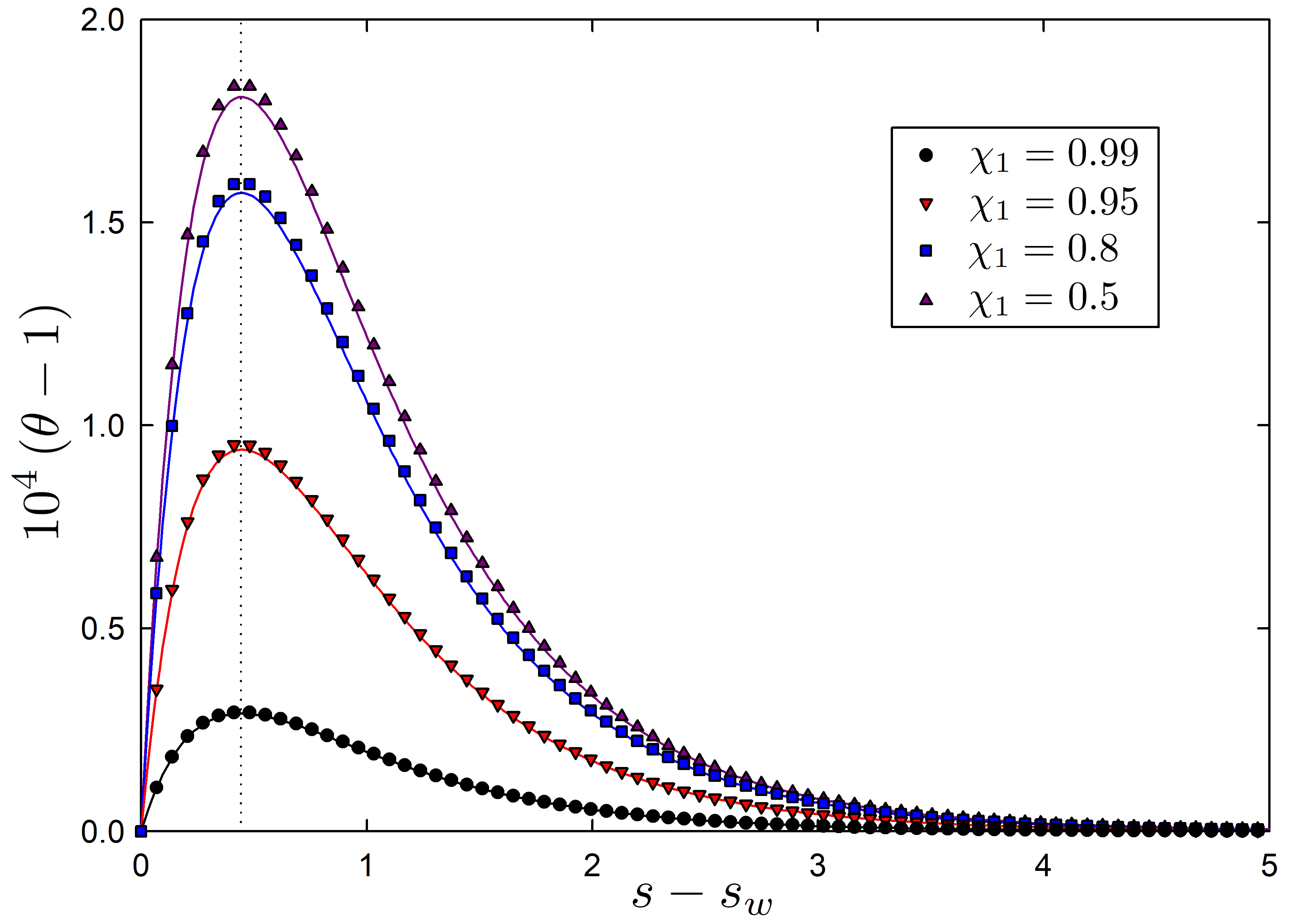}
\caption[]{(Color online) Kovacs hump in linear response. We have
  fixed the initial and final drivings, $\chi_{0}=1.05$ and $\chi=1$,
  and considered four values for the intermediate driving
  $\chi_1=\{0.5,0.8,0.95,0.99\}$. The linear response theory result
  \eqref{eq:Kovacs-factorisation-and-factors} (solid line) perfectly
  agrees with the numerical solution of the first Sonine approximation
  (symbols), Equations \eqref{eq:reduced-var-evol}. Also, the
  theoretical prediction for the maximum \eqref{eq:max-pos}, which
  again agrees with the numerical results, is plotted (dotted
  line). \label{fig3}}
\end{center}
\end{figure}

In order to check the above results, we have fixed the initial and
final drivings in the Kovacs protocol $\chi_{0}$ and $\chi$ and
changed the intermediate driving value $\chi_{1}$. We do so to
simplify the comparison, because the time scale $s$ involves the
steady value of the temperature, see Equation \eqref{eq:theta-A2-def}.
Note that the smaller $\chi_{1}$ is, the shorter the waiting time
becomes.  Therefore, one expects to get a Kovacs hump whose maximum
remains at $s-s_w\simeq 0.44$ but raises as $\chi_{1}$ decreases. This
is shown in Figure \ref{fig3}, where the numerical solution of the
first Sonine approximation equations \eqref{eq:reduced-var-evol} and
the analytical result \eqref{eq:Kovacs-factorisation-and-factors} are
compared.  Their agreement is almost perfect for the two lowest
curves, corresponding to $\chi_{1}=0.99$ and $\chi_{1}=0.95$, as
expected, but is still remarkably good for the two topmost ones,
corresponding to the not-so-small jumps for $\chi_{1}=0.8$ and
$\chi_{1}=0.5$.

\subsubsection{Nonlinear regime}

\begin{figure}
\begin{center}
\includegraphics[width=0.7 \textwidth]{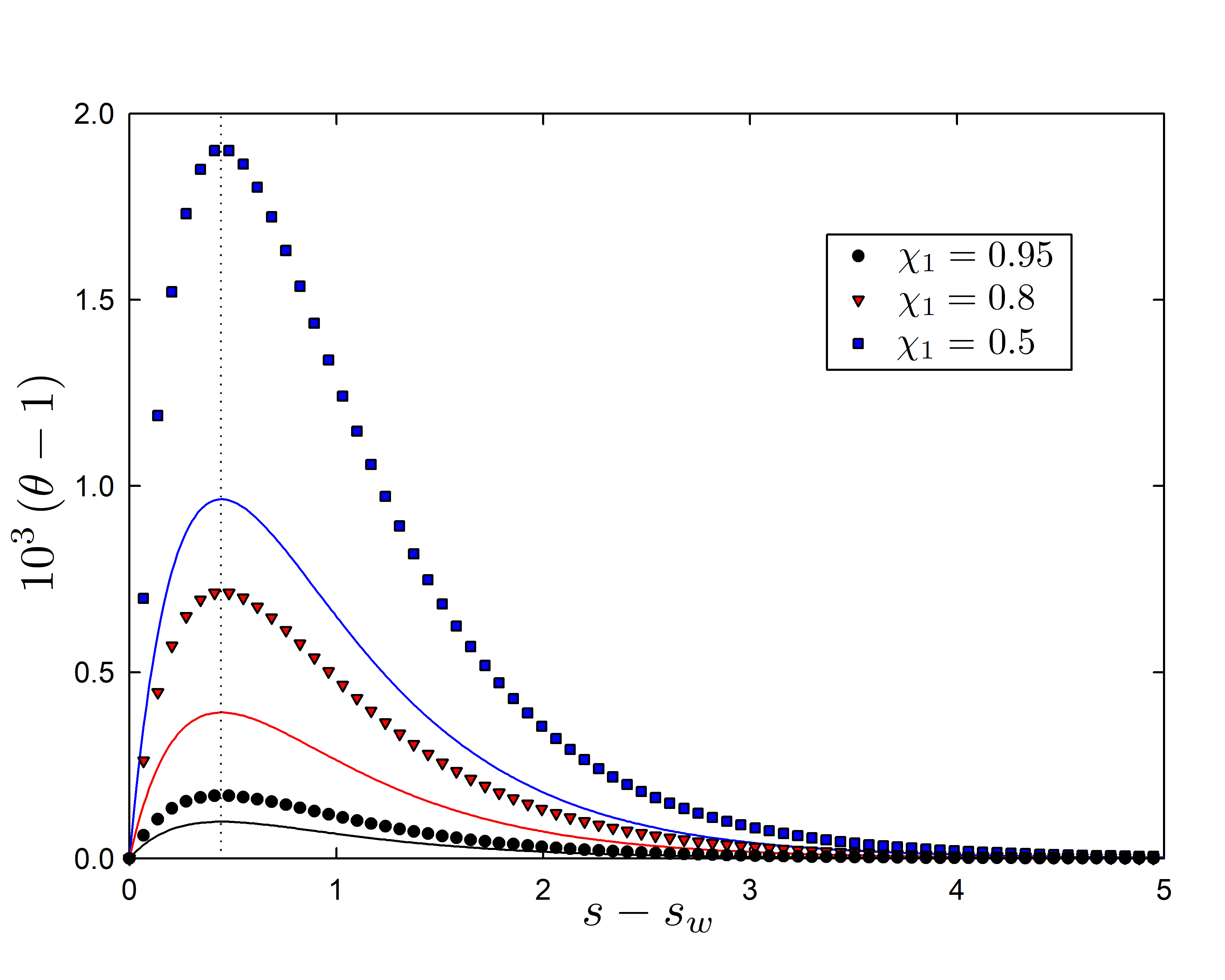}
\caption[]{(Color online) Kovacs hump out of the linear regime.  The
  initial driving is much higher than that in Figure \ref{fig3},
  $\chi_{0}=10$, whereas the final and intermediate values of the
  driving are again $\chi=1$ and $\chi_1=\{0.5,0.8,0.95\}$.  The
  linear response theoretical expression \eqref{eq:Kovacs-Sonine}
  (solid line) remains quite below the numerical solutions of the
  first Sonine approximation \eqref{eq:reduced-var-evol}
  (symbols). The theoretical expression for the maximum in linear
  response \eqref{eq:max-pos} (dotted line) still gives a good
  description thereof, see also Figure \ref{fig5}. \label{fig4}}
\end{center}
\end{figure}

Also we explore the Kovacs effect out of the linear
regime. Figure~\ref{fig4} is similar to Figure~\ref{fig3}, but for
larger temperature (or driving) jumps. We have also fixed the initial
and final values of the driving, $\chi_{0}=10$ and $\chi=1$. The
intermediate values of the driving are the same as in the linear case
except for the largest one, $\chi_{1}=0.99$, which we have omitted for
the sake of clarity (its hump is too small in the scale of the
figure).  Now, the linear response theory results just provide the
qualitative behaviour of the hump, correctly predicting the position
of the maximum but not its height. On the one hand, and consistently
with the numerical results in an active matter model
\cite{kursten_giant_2017}, the Kovacs hump out the linear response
regime is larger than the prediction of linear response theory.  On
the other hand, the position of the maximum remains basically
unchanged and its height still increases as $\chi_{1}$ decreases.

We also compare our analytical expansion in $a_{2}^{\ness}$ with the
numerical solutions of Equations \eqref{eq:reduced-var-evol} for large
jumps. Specifically, in order to make things as simple as possible, we
choose $\chi_{1}=0$. If the waiting time is long enough, the system
reaches the homogeneous cooling state and
$a_{2}(s_{w})=a_{2}^{\text{HCS}}$, which is then the initial condition
for the final stage of the Kovacs protocol. Moreover, we can compute
the location of the maximum of the hump from Equation
\eqref{eq:expansion_a2s-theta}, obtaining

\begin{equation}
\label{maximium_expansion}
s_{k}-s_{w}=\frac{2}{\kappa_{2}-2-\beta} \ln \left( \frac{2\kappa_{2}}{2+\beta}\right) \underset{\beta=1}{\simeq} 0.437.
\end{equation}
This result is numerically indistinguishable from that of linear
response, as given by Equation \eqref{eq:max-pos}, since the relative
error is around $1 \%$.

\begin{figure}
\begin{center}
\includegraphics[width=0.7 \textwidth]{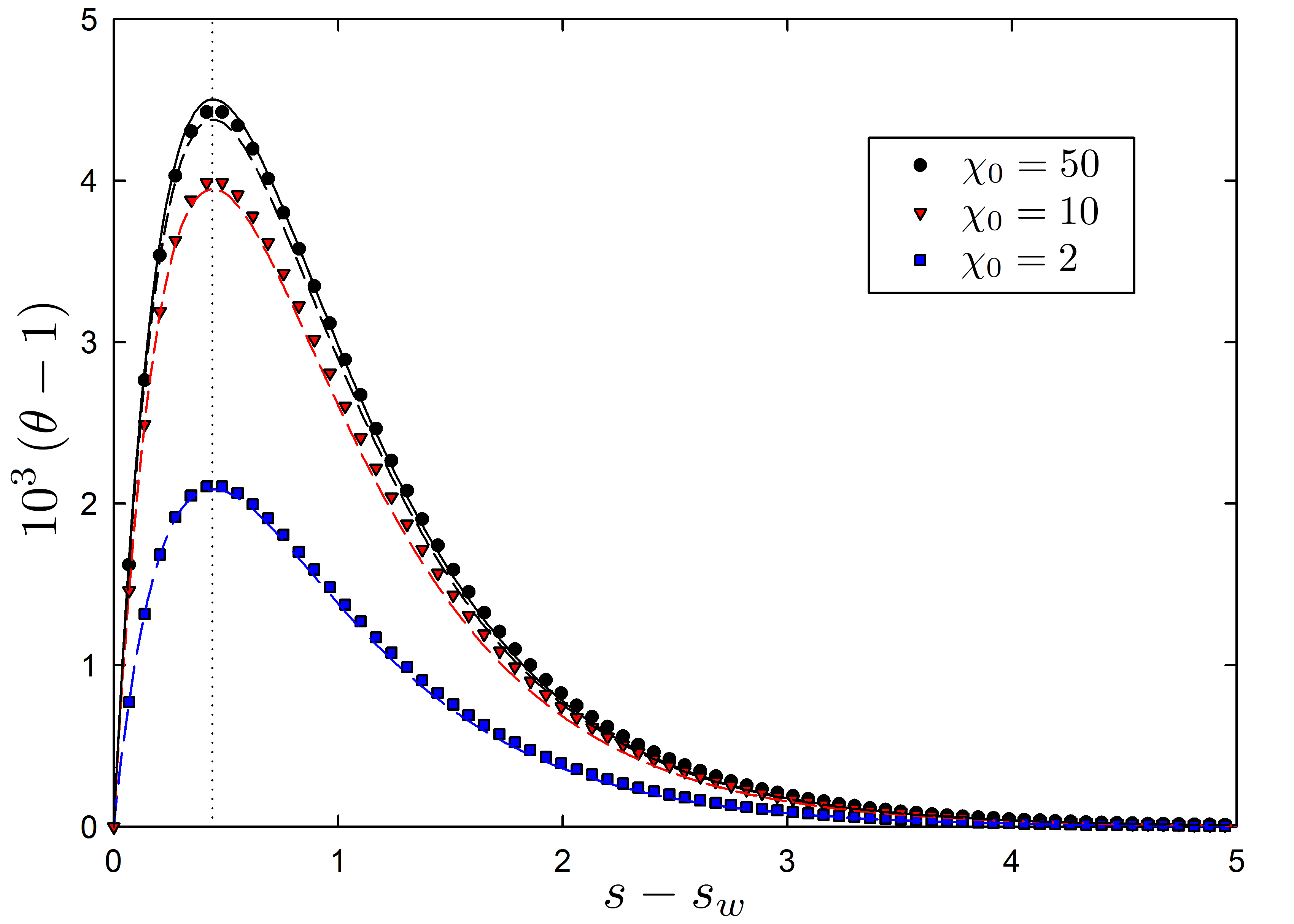}
\caption[]{(Color online) Kovacs hump in the nonlinear regime and
  prediction of the perturbative expansion in $a_{2}^{\ness}$. We have
  considered the following values of the drivings:
  $\chi_0=\{2,10,50\} $, $\chi_1=0$ and $\chi=1$. Symbols stand for
  the numerical solutions of the first Sonine approximation
  \eqref{eq:reduced-var-evol}, whereas lines correspond to the
  theoretical expression \eqref{eq:expansion_a2s-theta}. For the solid
  line, $a_{2}^{\text{ini}}=a_{2}^{\text{HCS}}$, while for the dashed
  lines we have used the value of $a_{2}^{\text{ini}}$ in the
  numerical solution. A perfect agreement is observed. Finally, we
  have plotted the theoretical expression for the maximum position in
  nonlinear response (dotted line), Equation
  \eqref{maximium_expansion}, which also shows an excellent agreement
  with numerics. \label{fig5}}
\end{center}
\end{figure}

In Figure \ref{fig5}, we put forward a comparison between the Kovacs
hump obtained from the numerical solution of the first Sonine
approximation equations and our theoretical expression for the
nonlinear regime, Equation \eqref{eq:expansion_a2s-theta}. Fixing
$\xi_{1}=0$ and $\xi=1$, as $\xi_{0}$ increases (as the waiting time
is increased), the hump approaches Equation \eqref{eq:expansion_a2s-theta}
with $a_{2}^{\text{ini}}=a_{2}^{\text{HCS}}$. Moreover, our theory
perfectly reproduces all the numerical curves when we substitute the
actual values of $a_{2}^{\text{ini}}$ into Equation
\eqref{eq:expansion_a2s-theta} .

\subsection{Monotonicity of an $H$-functional}

The non-monotonicity in the relaxation of the granular temperature
that is brought about by the Kovacs protocol is not automatically
transferred to other relevant physical magnitudes. Specifically, here
we deal with $H$-functional

\begin{equation}\label{eq:H-func}
H(t)=\int_{-\infty}^{+\infty}dv f(v,t) \log \left[\frac{f(v,t)}{f_{\ness}(v)} \right],
\end{equation}
There is strong numerical evidence about $H(t)$ being a Lyapunov
functional for granular fluids, thus allowing it to be considered as
an out-of-equilibrium entropy relative to that of the NESS, in this
context \cite{marconi_about_2013,de_soria_towards_2015}. Moreover, it
has been analytically proven that $H(t)$ is a Lyapunov functional in
our system for the Maxwell collision rule $\beta=0$
\cite{plata_global_2017}.

\begin{figure}
\begin{center}
\includegraphics[width=\textwidth]{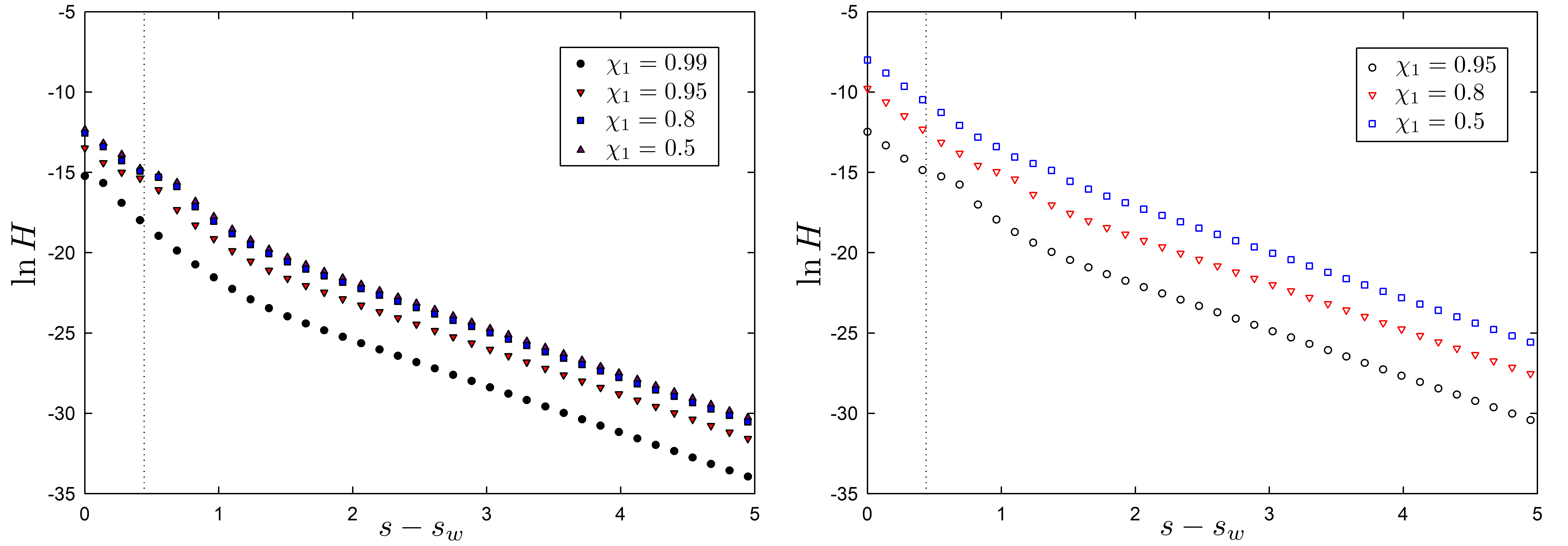}
\caption[]{(Color online) Time evolution of the $H$-functional. The
  relaxation of $H$ is shown to be monotonic even for Kovacs-like
  experiments. The left and right panels corresponds to the protocols
  in Figure \ref{fig3} (filled symbols) and \ref{fig4} (open symbols),
  that is, to the linear and nonlinear regimes. The vertical dotted
  line marks the theoretical position of the maximum in the
  corresponding regime. \label{fig6}}
\end{center}
\end{figure}

We have computed $H(t)$ numerically from Equation \eqref{eq:H-func}
within the first Sonine approximation,\footnote{That is, we have
  substituted both $f(v,t)$ and $f_{\ness}(v)$ by their expressions in
  the first Sonine approximation and calculated the integral
  numerically.} for the Kovacs protocols considered in Figures
\ref{fig3} and \ref{fig4}. Once more, we have taken $\beta=1$, for
which the analytical proof in \cite{plata_global_2017} does not
hold. The results are shown in Figure \ref{fig6} and make it clear
that $H(t)$ still monotonically decreases for the Kovacs-like
protocols, in both the linear (left panel) and nonlinear (right panel)
regimes. At the time of the maximum in the hump, $s-s_{w}\simeq 0.44$,
no special signature is observed in the entropy.

\section{Discussion}

One of the main results in our paper is the extension of the linear
response expression for the Kovacs hump in thermal systems, as given
by Equations \eqref{eq:kovacs-thermal-linear} and
\eqref{eq:T-tw-relation}, to the realm of athermal systems, Equations
\eqref{eq:evol-Yav-phi} and \eqref{eq:Kovacs-condition}.  This
extension is (i) not trivial, since athermal systems tend in the long
time limit to a NESS, not to an equilibrium state and (ii) very
general, since it can be done starting from the evolution equations at
either the mesoscopic or the macroscopic level of
description. Therefore, it means that a Kovacs-like effect is to be
expected for a very wide class of physical properties $Y$ in a very
wide class of systems, basically as long as the relaxation function
$\phi_{Y}(t)$ to the NESS is monotonic.

This theoretical result has been checked in a class of systems that
mimic the dynamics of the shear component of the velocity of a
granular fluid. In the linear response regime, we have found a good
agreement between the theoretical prediction and the numerical
results. Furthermore, we have investigated how the linear response
result extends to the nonlinear regime. In this region, the linear
response theory results remain useful at a qualitative level, but the
actual values for the hump lie well above the linear
prediction. Interestingly, this kind of \textit{giant} Kovacs hump has
been recently reported in active systems \cite{kursten_giant_2017}. In
addition, the nonlinear Kovacs hump can be theoretically explained by
an expansion in the excess kurtosis.

The work presented here also opens new interesting perspectives for
future research. First, our analysis of the Kovacs effect in the
model, being restricted to the quasi-elastic limit, has not found the
\textit{anomalous} behaviour shown by a gas of inelastic hard spheres
for large enough inelasticity in the nonlinear regime
\cite{prados_kovacs-like_2014,trizac_memory_2014}. The possibility of
such a behaviour in linear response, either in the model or in the
granular gas, deserves further investigation.  Second, our work
clearly shows the compatibility of the non-monotonic decay of the
granular temperature (or the corresponding relevant physical variable)
and the monotonic decay of the nonequilibrium entropy or
$H$-functional
\cite{marconi_about_2013,de_soria_towards_2015,plata_global_2017,brey_normal_1993}.
In this respect, to elucidate if the hump leaves some signature in the
decay of the nonequilibrium entropy is compelling.

\vspace{6pt} 


\acknowledgments{We acknowledge the support of the Spanish Ministerio
  de Economía y Competitividad through Grant FIS2014-53808-P. Carlos A. Plata also acknowledges the support from the FPU Fellowship Programme of the Spanish Ministerio de Educación, Cultura y Deporte through Grant FPU14/00241.}

\authorcontributions{All authors contributed equally to this work. All authors have read and approved the final manuscript.}

\conflictsofinterest{The authors declare no conflict of interest. The founding sponsors had no role in the design of the study; in the collection, analyses, or interpretation of data; in the writing of the manuscript, and in the decision to publish the results.} 



\appendixtitles{yes} 
\appendixsections{one} 
\appendix
\section{Simulation algorithm}
\label{appA}
We have made use of a residence time algorithm that gives the
numerical integration of a master equation in the limit of
infinite trajectories \cite{bortz_new_1975,prados_dynamical_1997}. The
basic numerical recipe is as follows:
\begin{enumerate}

\item\label{stepone} At time $\tau$, a random ``free time''
  $\tau_{f} > 0$ is extracted with an exponential probability density
  $\Omega(\bm{v})\exp[-\Omega(\bm{v})\tau_{f}]$, where
  $\Omega(\bm{v})=\sum_l \omega |v_l-v_{l+1}|^\beta$ depends on the
  state of the system $\bm{v}$;

\item Time is advanced by such a free time
$\tau \to \tau+\tau_f$;

\item A pair $(l,l+1)$ is chosen to collide with probability
  $\omega|v_l-v_{l+1}|^\beta /\Omega(\bm{v})$;

\item All particles are heated by the stochastic thermostat, by adding
   independent  Gaussian random numbers of zero mean and variance $\chi\tau_{f}$ to their velocities;

 \item In order to conserve momentum, the mean value of the random
   numbers generated in the previous step is subtracted from the
   velocities of all particles;

\item The process is repeated from step~\ref{stepone}.
\end{enumerate}




\externalbibliography{yes}
\bibliography{Mi-biblioteca-18-sep-2017,Granular-v2}

\begin{thebibliography}{-------}
\providecommand{\natexlab}[1]{#1}

\bibitem[Kovacs \em{et~al.}(1979)Kovacs, Aklonis, Hutchinson, and
  Ramos]{kovacs_isobaric_1979}
Kovacs, A.J.; Aklonis, J.J.; Hutchinson, J.M.; Ramos, A.R.
\newblock Isobaric volume and enthalpy recovery of glasses. {II}. {A}
  transparent multiparameter theory.
\newblock {\em Journal of Polymer Science: Polymer Physics Edition} {\bf 1979},
  {\em 17},~1097--1162.

\bibitem[Bertin \em{et~al.}(2003)Bertin, Bouchaud, Drouffe, and
  Godreche]{bertin_kovacs_2003}
Bertin, E.M.; Bouchaud, J.P.; Drouffe, J.M.; Godreche, C.
\newblock The {Kovacs} effect in model glasses.
\newblock {\em Journal of Physics A: Mathematical and General} {\bf 2003}, {\em
  36},~10701.

\bibitem[Buhot(2003)]{buhot_kovacs_2003}
Buhot, A.
\newblock Kovacs effect and fluctuation–dissipation relations in 1D
  kinetically constrained models.
\newblock {\em Journal of Physics A: Mathematical and General} {\bf 2003}, {\em
  36},~12367.

\bibitem[Mossa and Sciortino(2004)]{mossa_crossover_2004}
Mossa, S.; Sciortino, F.
\newblock Crossover (or {Kovacs}) effect in an aging molecular liquid.
\newblock {\em Phys. Rev. Lett.} {\bf 2004}, {\em 92},~045504.

\bibitem[Aquino \em{et~al.}(2006)Aquino, Leuzzi, and
  Nieuwenhuizen]{aquino_kovacs_2006}
Aquino, G.; Leuzzi, L.; Nieuwenhuizen, T.M.
\newblock Kovacs effect in a model for a fragile glass.
\newblock {\em Phys. Rev. B} {\bf 2006}, {\em 73},~094205.

\bibitem[Prados and Brey(2010)]{prados_kovacs_2010}
Prados, A.; Brey, J.J.
\newblock The {Kovacs} effect: a master equation analysis.
\newblock {\em J. Stat. Mech.} {\bf 2010}, p. P02009.

\bibitem[Diezemann and Heuer(2011)]{diezemann_memory_2011}
Diezemann, G.; Heuer, A.
\newblock Memory effects in the relaxation of the {Gaussian} trap model.
\newblock {\em Physical Review E} {\bf 2011}, {\em 83},~031505.

\bibitem[Ruiz-García and Prados(2014)]{ruiz-garcia_kovacs_2014}
Ruiz-García, M.; Prados, A.
\newblock Kovacs effect in the one-dimensional {Ising} model: {A} linear
  response analysis.
\newblock {\em Phys. Rev. E} {\bf 2014}, {\em 89}.

\bibitem[Van~Kampen(1992)]{van_kampen_stochastic_1992}
Van~Kampen, N.G.
\newblock {\em Stochastic processes in Physics and Chemistry}; North-Holland,
  1992.

\bibitem[P\"oschel and Luding(2001)]{PL01}
P\"oschel, T.; Luding, S., Eds.
\newblock {\em Granular Gases}; Vol. 564, {\em Lecture Notes in Physics},
  Springer: Berlin,  2001.

\bibitem[Van~Noije and Ernst(1998)]{van_noije_velocity_1998}
Van~Noije, T.P.C.; Ernst, M.H.
\newblock Velocity distributions in homogeneous granular fluids: the free and
  the heated case.
\newblock {\em Granul. Matter} {\bf 1998}, {\em 1},~57--64.

\bibitem[Prados and Trizac(2014)]{prados_kovacs-like_2014}
Prados, A.; Trizac, E.
\newblock Kovacs-{Like} {Memory} {Effect} in {Driven} {Granular} {Gases}.
\newblock {\em Phys. Rev. Lett.} {\bf 2014}, {\em 112},~198001.

\bibitem[Trizac and Prados(2014)]{trizac_memory_2014}
Trizac, E.; Prados, A.
\newblock Memory effect in uniformly heated granular gases.
\newblock {\em Phys. Rev. E} {\bf 2014}, {\em 90},~012204.

\bibitem[Lahini \em{et~al.}(2017)Lahini, Gottesman, Amir, and
  Rubinstein]{lahini_nonmonotonic_2017}
Lahini, Y.; Gottesman, O.; Amir, A.; Rubinstein, S.M.
\newblock Nonmonotonic {Aging} and {Memory} {Retention} in {Disordered}
  {Mechanical} {Systems}.
\newblock {\em Physical Review Letters} {\bf 2017}, {\em 118},~085501.

\bibitem[Kürsten \em{et~al.}(2017)Kürsten, Sushkov, and
  Ihle]{kursten_giant_2017}
Kürsten, R.; Sushkov, V.; Ihle, T.
\newblock Giant {Kovacs}-{Like} {Memory} {Effect} for {Active} {Particles}.
\newblock {\em arXiv:1705.05275} {\bf 2017}.

\bibitem[Lasanta \em{et~al.}(2015)Lasanta, Manacorda, Prados, and
  Puglisi]{lasanta_fluctuating_2015}
Lasanta, A.; Manacorda, A.; Prados, A.; Puglisi, A.
\newblock Fluctuating hydrodynamics and mesoscopic effects of spatial
  correlations in dissipative systems with conserved momentum.
\newblock {\em New J. Phys.} {\bf 2015}, {\em 17},~083039.

\bibitem[Manacorda \em{et~al.}(2016)Manacorda, Plata, Lasanta, Puglisi, and
  Prados]{manacorda_lattice_2016}
Manacorda, A.; Plata, C.A.; Lasanta, A.; Puglisi, A.; Prados, A.
\newblock Lattice {Models} for {Granular}-{Like} {Velocity} {Fields}:
  {Hydrodynamic} {Description}.
\newblock {\em J. Stat. Phys.} {\bf 2016}, {\em 164},~810--841.

\bibitem[Plata \em{et~al.}(2016)Plata, Manacorda, Lasanta, Puglisi, and
  Prados]{plata_lattice_2016}
Plata, C.A.; Manacorda, A.; Lasanta, A.; Puglisi, A.; Prados, A.
\newblock Lattice models for granular-like velocity fields: finite-size
  effects.
\newblock {\em J. Stat. Mech.} {\bf 2016}, p. 093203.

\bibitem[Marconi \em{et~al.}(2013)Marconi, Puglisi, and
  Vulpiani]{marconi_about_2013}
Marconi, U.M.B.; Puglisi, A.; Vulpiani, A.
\newblock About an {H}-theorem for systems with non-conservative interactions.
\newblock {\em J. Stat. Mech.} {\bf 2013}, p. P08003.

\bibitem[García~de Soria \em{et~al.}(2015)García~de Soria, Maynar, Mischler,
  Mouhot, Rey, and Trizac]{de_soria_towards_2015}
García~de Soria, M.I.; Maynar, P.; Mischler, S.; Mouhot, C.; Rey, T.; Trizac,
  E.
\newblock Towards an {H}-theorem for granular gases.
\newblock {\em J. Stat. Mech.} {\bf 2015}, p. P11009.

\bibitem[Plata and Prados(2017)]{plata_global_2017}
Plata, C.A.; Prados, A.
\newblock Global stability and $H$-theorem in lattice models with
  nonconservative interactions.
\newblock {\em Physical Review E} {\bf 2017}, {\em 95},~052121.

\bibitem[Brey and Prados(1993)]{brey_stretched_1993}
Brey, J.J.; Prados, A.
\newblock Stretched exponential decay at intermediate times in the
  one-dimentional {Ising} model at low temperatures.
\newblock {\em Physica A} {\bf 1993}, {\em 197},~569--582.

\bibitem[Prasad \em{et~al.}(2017)Prasad, Sabhapandit, Dhar, and
  Narayan]{prasad_driven_2016}
Prasad, V.V.; Sabhapandit, S.; Dhar, A.; Narayan, O.
\newblock Driven inelastic {Maxwell} gas in one dimension.
\newblock {\em Phys. Rev. E} {\bf 2017}, {\em 95},~022115.

\bibitem[Ben-Naim and Krapivsky(2003)]{BK03}
Ben-Naim, E.; Krapivsky, P.L.
\newblock The inelastic {Maxwell} model.
\newblock  Granular Gas Dynamics; P\"oschel, T.; Brilliantov, N., Eds.;
  Springer: Berlin,  2003; Vol. 624, {\em Lecture Notes in Physics}, pp.
  65--94.

\bibitem[Ernst \em{et~al.}(2006)Ernst, Trizac, and Barrat]{ETB06a}
Ernst, M.H.; Trizac, E.; Barrat, A.
\newblock The rich behavior of the Boltzmann equation for dissipative gases.
\newblock {\em EPL} {\bf 2006}, {\em 76},~56--62.

\bibitem[Montanero and Santos(2000)]{MS00}
Montanero, J.M.; Santos, A.
\newblock Computer simulation of uniformly heated granular fluids.
\newblock {\em Granul. Matter} {\bf 2000}, {\em 2},~53--64.

\bibitem[Maynar \em{et~al.}(2009)Maynar, García~de Soria, and
  Trizac]{maynar_fluctuating_2009}
Maynar, P.; García~de Soria, M.I.; Trizac, E.
\newblock Fluctuating hydrodynamics for driven granular gases.
\newblock {\em Eur. Phys. J. Spec. Top.} {\bf 2009}, {\em 179},~123--139.

\bibitem[Prasad \em{et~al.}(2013)Prasad, Sabhapandit, and
  Dhar]{prasad_high-energy_2013}
Prasad, V.V.; Sabhapandit, S.; Dhar, A.
\newblock High-energy tail of the velocity distribution of driven inelastic
  {Maxwell} gases.
\newblock {\em EPL} {\bf 2013}, {\em 104},~54003.

\bibitem[Plata and Prados()]{PlataPradosunpub17}
Plata, C.A.; Prados, A.
\newblock (unpublished).

\bibitem[Abramowitz and Stegun(1972)]{AS72}
Abramowitz, M.; Stegun, I.A., Eds.
\newblock {\em Handbook of Mathematical Functions}, New York,  1972. Dover.

\bibitem[Brey and Prados(1993)]{brey_normal_1993}
Brey, J.J.; Prados, A.
\newblock Normal solutions for master equations with time-dependent transition
  rates: {Application} to heating processes.
\newblock {\em Phys. Rev. E} {\bf 1993}, {\em 47},~1541.

\bibitem[Bortz \em{et~al.}(1975)Bortz, Kalos, and Lebowitz]{bortz_new_1975}
Bortz, A.B.; Kalos, M.H.; Lebowitz, J.L.
\newblock A new algorithm for {Monte} {Carlo} simulation of {Ising} spin
  systems.
\newblock {\em J. Comput. Phys.} {\bf 1975}, {\em 17},~10--18.

\bibitem[Prados \em{et~al.}(1997)Prados, Brey, and
  Sánchez-Rey]{prados_dynamical_1997}
Prados, A.; Brey, J.J.; Sánchez-Rey, B.
\newblock A dynamical Monte Carlo algorithm for master equations with
  time-dependent transition rates.
\newblock {\em J. Stat. Phys.} {\bf 1997}, {\em 89},~709--734.

\end{thebibliography}


\end{document}